\input harvmac.tex


\Title{\vbox{\baselineskip12pt\hbox{hep-th/0001200}
\hbox{CALT-68-2260}\hbox{CITUSC/00-008}}}
{\vbox{
\centerline{D-branes on Orbifolds with Discrete Torsion }
\vskip 10pt
\centerline{And Topological Obstruction}
}}
\centerline{Jaume Gomis}
\medskip
\medskip
\medskip
\medskip
\centerline{\it Department of Physics}
\centerline{\it California Institute of Technology}
\centerline{\it Pasadena, CA 91125}
\centerline{\it and}
\centerline{\it Caltech-USC Center for Theoretical Physics} 
\centerline{\it University of Southern California}
\centerline{Los Angeles, CA 90089}

\medskip
\medskip
\noindent

We find the orbifold analog of the topological relation recently
found by Freed and Witten which restricts the allowed D-brane configurations
of Type II vacua with  a topologically non-trivial
flat $B$-field. The result relies in Douglas proposal -- which we
derive from worldsheet consistency conditions -- of embedding
projective representations on open string Chan-Paton factors 
when considering orbifolds with discrete torsion. The orbifold action 
on open strings
gives a natural definition of the algebraic K-theory group -- using twisted
cross products -- 
responsible for measuring Ramond-Ramond charges in orbifolds with
discrete torsion. We show that the correspondence between 
fractional
branes and Ramond-Ramond fields follows in an interesting fashion
from the way that discrete torsion is
implemented on open  and closed strings.

\smallskip

\Date{January 2000}

\newsec{Introduction, Results and Conclusions}

Orbifolds 
\nref\dhvw{L.Dixon, J. Harvey, C. Vafa and E.Witten, ``Strings on
Orbifolds I and II'', Nucl .Phys. {\bf B261} (1985) 678 
and Nucl. Phys. {\bf B274} (1986) 285.}%
\dhvw\ in string theory provide a tractable arena where CFT
can be used to describe perturbative vacua. In 
the more conventional geometric compactifications,
geometry and topology provide powerful techniques in
describing the long wavelength approximation to string theory. In a
sense, geometric compactifications and orbifolds provide a sort of 
dual description. In the latter,  CFT techniques are available
but topology is less manifest. On the other hand, conventional Calabi-Yau
compactification 
\nref\chsw{P. Candelas, G.T. Horowitz, A. Strominger and E. Witten,
``Vacuum Configurations for Superstrings'', Nucl. Phys. {\bf B258}
(1985) 46.}%
\chsw\ lacks an exact CFT formulation but a rich 
mathematical apparatus aids the analysis of 
the corresponding
supergravity approximation.

Perhaps surprisingly, orbifold CFT
\nref\dfms{L. Dixon, D. Friedan, E. Martinec and S.H. Shenker, ``The
Conformal Field Theory of Orbifolds'', Nucl. Phys. {\bf B282} (1997)
13.}%
\dhvw\dfms\
 seems to be able to realize  topological
relations satisfied by geometric compactifications
from worldsheet consistency conditions.
 A nice
example of this phenomenon is the interpretation
\nref\vafa{C. Vafa, ``Modular
 Invariance and Discrete Torsion on
Orbifolds'', Nucl. Phys. {\bf B273} (1986) 592.}%
\vafa\
of a restriction
imposed by modular invariance 
as the analog of the 
topological constraint requiring the space-time manifold to
have a vanishing  second Stieffel-Whitney class.

The original motivation of this work was to find the analog of
the topological formula recently found by Freed and
Witten\foot{Several aspects of this topological relation had already
been considered by Witten in
\nref\wittena{E. Witten, ``Baryons And Branes In Anti-de Sitter
Space'', JHEP {\bf 9807}:006, 1998, hep-th/9805112.}%
\nref\wittenb{E. Witten, ``D-Branes And K-Theory'',  JHEP {\bf
9812}:025, 1998, hep-th/9810188.}%
\wittena\wittenb .} 
\nref\freedwitten{D.S. Freed and E. Witten, ``Anomalies in String
Theory with D-branes'', hep-th/9907189.}%
\freedwitten\ 
 in the context of orbifolds. 
Let us briefly explain
their results in a language that will be convenient for what follows.
Given a space-time
manifold $X$ and a submanifold $Y\subset X$, their formula constraints the
configuration of 
D-branes allowed to wrap $Y$.
A
careful analysis \freedwitten\ of string worldsheet global anomalies 
in the presence of D-branes and a flat Neveu-Schwarz B-field
-- so that the curvature $H=dB$ is zero -- imposes,
 for a class of backgrounds,
the following topological relation\foot{This formula can have an
additional term that depends   on the topology of $Y$. We will ignore
this correction since it vanishes when the 
second 
Stieffel-Whitney class of $Y$ is trivial and the orbifold model
satisfies a relation which can be identified with the vanishing of this class.}
\eqn\bfield{
i^*[H]=0,}
where $[H]\in H^3(X,Z)$ determines the topological class of the
B-field and $i^*[H]$ is the restriction of $[H]$ to the D-brane
worldvolume $Y\subset X$. Since $[i^*[H]]\in  H^3(Y,Z)$ is a torsion
class in cohomology, so
 that there is a smallest non-zero integer $m$ such that 
$m\cdot[i^*[H]]\simeq 0$, the anomaly relation \bfield\ can only be
satisfied whenever there are a multiple of $m$ D-branes wrapping $Y\subset X$.
Therefore, in a background with a topologically non-trivial flat 
B-field such that
$m\cdot[i^*[H]]\simeq 0$, the  charge of the minimal  D-brane
configuration wrapping $Y$ is $m$ times
bigger than in a background with trivial B-field.
The anomaly relation \bfield\ expresses in topological terms that
D-branes in string theory are not pure geometric constructs, but
that the allowed configurations may depend on discrete choices -- like
the choice of $[H]$ -- of the
string background. In some cases, this fact is realized by the K-theory
\nref\minmo{R. Minasian and G. Moore, `` K-theory and Ramond-Ramond
charge'', JHEP {\bf 9711} (1997) 002.}%
classification \minmo\wittenb\ of Ramond-Ramond charges\foot{The work
of Sen describing D-branes via tachyon condensation of unstable systems
-- see for example
\nref\sen{A. Sen, `` Stable Non-BPS States in String Theory'', JHEP
{\bf 9806} (1998) 007; ``Stable Non-BPS Bound States of BPS
D-branes'',  JHEP {\bf 9808} (1998) 010; ``Tachyon 
Condensation on the Brane Antibrane System'', JHEP {\bf 9808} (1998)
012; ``$SO(32)$ Spinors of Type I and Other Solitons on
Brane-Antibrane Pair'', JHEP {\bf 9809} (1998) 023.}%
\nref\berga{O. Bergman and M.R. Gaberdiel, ``Stable non-BPS
D-particles'', Phys. Lett. {\bf B441} (1998) 133, hep-th/9806155.}%
\nref\fgls{M. Frau, L. Gallot, A. Lerda and P. Strigazzi, ``Stable
non-BPS D-branes in Type I string theory'', hep-th/9903123.}%
\sen\berga\fgls -- was crucial in making the identification between
D-branes and K-theory. See 
\nref\horava{P. Ho\v rava, ``Type IIA D-Branes, K-Theory, and Matrix
Theory'', Adv. Theor. Math. Phys. {\bf 2} (1999) 1373,
hep-th/9812135.}%
\horava\ for a Type IIA discussion.}.
For example, whenever $[H]$ is not trivial, Type IIB
D-brane charges are classified by the twisted topological K-theory group
$\hbox{K}_{[H]}(X)$
\nref\kap{A. Kapustin, ``D-branes in a topologically nontrivial
B-field'', hep-th/9909089.}%
\wittenb\kap\
instead of the more
conventional group $\hbox{K}(X)$ used whenever $[H]$
is trivial.

In this note we find the analogous phenomenon for orbifold models.
 Given a compact orbifold $T^6/\Gamma$ with discrete
torsion\foot{The simplest supersymmetric orbifold model with discrete
 torsion appears when the orbifold is three complex dimensional. For
 concreteness, we will study in section $3$ the case $\Gamma=Z_n\times
 Z_{n^{\prime}}$. In 
\nref\fiq{A. Font, L.E. Ib\'a\~nez and F. Quevedo, ``$Z_n\times
 Z_{n^{\prime}}$ Orbifolds and Discrete Torsion'', Phys. Lett. {\bf
 B217} (1989) 272.}
\fiq , the values of $n$ and ${n^{\prime}}$ so that $\Gamma$ acts
 cristallographically were found.}
 \vafa , 
we show that the charge of the minimal D6-brane configuration wrapping 
 the 
orbifold is an integer multiple bigger than the minimal charge when one
considers conventional orbifolds (without discrete torsion). This
 result  parallels  the consequences that stem from \bfield . 
Roughly speaking, turning on
discrete torsion in the orbifold corresponds to turning on a flat
 topologically non-trivial 
$B$-field in a geometric  compactification and the
conventional orbifold corresponds to the case where the $B$-field is
trivial. This suggest that discrete torsion in string theory is
 intimately related to torsion
 in homology $[15\hskip-3pt-\hskip-3pt 19]$\foot{In many examples the
 relation is not direct and the 
 correspondence between torsion in homology and discrete torsion is
 only visible after an irrelevant perturbation,
\nref\vafawitten{C. Vafa and E. Witten, ``On Orbifolds with Discrete
Torsion'', J. Geom. Phys. {\bf 15} (1995) 189, hep-th/9409188.}%
\nref\asmorb{P.S. Aspinwall and D.R. Morrison, ``Chiral Rings Do Not
Suffice: $N=(2,2)$ Theories with Nonzero Fundamental Group'',
Phys. Lett. {\bf B334} (1994) 79, hep-th/9406032.}%
\nref\asmor{P.S. Aspinwall and D.R. Morrison, ``Stable Singularities
in String Theory'', Commun. Math. Phys. {\bf 178} (1996) 115, hep-th/9503208.}%
\nref\sharpe{E.R. Sharpe, ``Discrete Torsion and Gerbes I and II'',
hep-th/9909108 and hep-th/9909120.}%
\nref\leighbe{D. Berenstein and R.G. Leigh, ``Discrete Torsion,
AdS/CFT and duality'',  hep-th/0001055.}%
as in for example \asmor . }.

A crucial ingredient in deriving this result is a careful treatment of
open strings in orbifolds with discrete torsion.
Douglas 
\nref\mike{M.R. Douglas, ``D-branes and Discrete Torsion'',
hep-th/9807235.}%
\mike\ has proposed that discrete torsion should be  implemented on open
strings by embedding a projective representation of the orbifold
group on Chan-Paton factors. Whether $\Gamma$ admits discrete torsion
and projective representations depends on its cohomology via
$H^2(\Gamma,U(1))$. This alone suggest the correlation between
discrete torsion and projective representations.
We show that worldsheet
consistency conditions uniquely determine the action of the orbifold
group on Chan-Paton factors once a closed string orbifold model is
specified. This result can be derived by demanding that open and closed
strings interact properly in the orbifold, so that the orbifold
group $\Gamma$ is conserved by their interactions.
A careful account of what  is discrete torsion is essential in this
derivation and we shall present its description in section $2$.

The effects of discrete torsion can be incorporated to define a
 K-theory group which measures Ramond-Ramond
charges in orbifolds with discrete torsion. 
 In
the algebraic\foot{See 
\nref\gupe{S. Gukov and V. Periwal, ``Dbrane Phase Transitions and
 Monodromy in K-theory'', hep-th/9908166.}%
\gupe\ for prior use of the algebraic approach to K-theory in string theory.}
 approach to equivariant K-theory via cross products
\nref\black{B. Blackadar, ``K-Theory for Operator Algebras'',
 Springer-Verlag.}%
\black ,
one incorporates discrete torsion by twisting
 the cross product by a
 cocycle\foot{The multiplication law of the cross product $C(X)\propto
 \Gamma$, where $C(X)$ is the algebra of continous functions on $X$, 
 is twisted by a cocycle $c\in H^2(\Gamma,U(1))$ and defines and associative
 group ring generated by the elements of $\Gamma$ with $C(X)$
 coeffitients. The Grothendieck group of the twisted crossed product
 yields the K-theory group  $\hbox{K}_\Gamma^{[c]}(X)$. See \black\
 for a more complete discussion.}
 $c\in H^2(\Gamma,U(1))$ corresponding to 
the choice of
 discrete torsion and projective representation. It turns out that
whenever the action of a finite group $\Gamma$ on $X$ 
is free, the algebraic K-theory of twisted cross
 products $\hbox{K}_\Gamma^{[c]}(X)$ is 
isomorphic 
\nref\rae\will{I. Raeburn and D.P. Williams, ``Pull-backs of
 $C^*$-Algebras and Crossed Products by Diagonal Actions'', Trans. AMS
 {\bf 287} (1985) 755.}%
\rae\ to the twisted topological  K-theory group
 $\hbox{K}_{[H]}(X/\Gamma)$ which classifies D-branes in a background with
topologically non-trivial flat B-field. The use of projective
 representations -- and therefore of cocycles -- provides a definition
 of K-theory which is the orbifold generalization of $\hbox{K}_{[H]}(X)$.

We show that the minimal D-brane charge for six-branes wrapping $T^6/\Gamma$ 
is larger for orbifolds with discrete torsion than for
conventional orbifolds by explicitly computing the D-brane
charge. The D6-brane charge can be extracted from a disk amplitude
with an insertion of the corresponding untwisted Ramond-Ramond vertex
operator. The same result can be obtained as in 
\nref\jaume{D.E. Diaconescu and J. Gomis, ``Fractional Branes and
Boundary States in Orbifold Theories'', hep-th/9906242.}%
\jaume\ using the  boundary state
formalism 
\nref\bsta{J.
Polchinski and Y. Cai, "Consistency of Open Superstring Theories",
Nucl. Phys. {\bf B296} (1988) 91;\parskip=0pt
\item{}
C. Callan, C. Lovelace, C. Nappi and S. Yost, "Loop Corrections to
Superstring Equations of  Motion", Nucl. Phys. {\bf B308}  (1988)
221;\parskip=0pt
\item{}
T. Onogi and N. Ishibashi,
 ``Conformal Field Theories On Surfaces With Boundaries And
 Crosscaps",
Mod. Phys. Lett. {\bf A4} (1989) 161;
\parskip=0pt
\item{}
N. Ishibashi,
 ``The Boundary And Crosscap States In Conformal Field Theories",
Mod. Phys. Lett. {\bf A4} (1989) 251.}%
\bsta . As shown in 
\nref\fract{D.E. Diaconescu, M.R. Douglas and J. Gomis, ``Fractional
 Branes and Wrapped Branes'', JHEP {\bf 9802} (1998) 013, hep-th/9712230.}%
\fract\jaume , the 
properly normalized untwisted Ramond-Ramond six-brane charge is
given by
\eqn\charge{
Q={d_R\over |\Gamma|},}
where $d_R$ is the dimension of the $R$-representation\foot{In the
more conventional setup of branes transverse to a non-compact orbifold, a bulk
brane is described by the regular representation so that $Q=1$ 
and a brane stuck at the singularity by an irreducible representation
and carries fractional untwisted charge \fract .}
 of $\Gamma$ acting on
the Chan-Paton factors and $|\Gamma|$ is the order of the group.
 The
minimal charge is therefore obtained by taking the smallest
irreducible representation of $\Gamma$. For open strings in
conventional orbifolds one must use standard (vectorial)
representations of $\Gamma$. For any discrete group $\Gamma$, the
smallest irreducible vectorial representation 
is always one-dimensional\foot{One
always has the trivial representation where each element $g_i\in
\Gamma$ is representated by $1$.}. As shown in section $2$, particular
projective representations must be used when dealing with orbifolds
with discrete torsion. A simple and important property of projective
representations is that there are no non-trivial\foot{Any group
$\Gamma$
 can have projective
representations. The important issue is whether a given projective
representation can be redefined to become a vectorial one. This will
become more clear in section $2$.}
one-dimensional projective
representations. Therefore, given a model with orbifold group $\Gamma$
and non-trivial $H^2(\Gamma,U(1))$, the charge of the minimal 
D6-brane configuration 
 when discrete torsion is turned on is given in terms of the
charge of the minimal D6-brane configuration 
when discrete torsion is turned off by
\eqn\ratio{
Q_{dis.tors.}=d^{proj}_RQ_{convent.}.}
$d^{proj}_R>1$ is dimension of the smallest irreducible projective
representation of $\Gamma$. Thus, the charge of a D6-brane wrapping
the entire orbifold is always larger when one consider orbifolds with
discrete torsion than when one considers conventional orbifolds.

The correlation between discrete torsion in the closed string sector
and the use of projective representations on open strings provides
a natural description of fractional branes 
\nref\miked{M.R. Douglas, ``Enhanced Gauge Symmetry in M(atrix)
Theory'', JHEP {\bf 9707} (1997) 004, hep-th/9612126.}%
\miked\fract\ 
in these models. For simplicity, let's consider D0-branes sitting at a
point in a conventional non-compact orbifold $C^3/\Gamma$. The 
charge vector of any 
zero-brane state  lies 
in a charge lattice generated by a basis of charge vectors.
Each irreducible
representation of $\Gamma$ is associated with  a basis vector of the charge
lattice. This result can be shown both from CFT 
\fract\jaume\ and the K-theory approach 
\minmo\wittenb\ to D-brane
charges using equivariant K-theory
\nref\compe{H. Garcia-Compean, ``D-branes in Orbifold Singularities
and Equivariant K-Theory'', Nucl. Phys. {\bf B557} (1999) 480,
hep-th/9812226.}%
\nref\sergi{S. Gukov, ``K-Theory, Reality, and Orientifolds'',
hep-th/9901042.}%
\wittenb\compe\sergi . The closed string spectrum yields
a massless Ramond-Ramond one-form potential 
for each twisted sector, but there are as
many twisted sectors as irreducible representations of
$\Gamma$, so that indeed one can associate a generator of the charge
lattice with each  irreducible representation. A particular
zero-brane state is uniquely specified by the choice of representation
of $\Gamma$ on its Chan-Paton factors, but any representation of
$\Gamma$ can be uniquely decomposed into a particular sum of its irreducible
ones. Therefore, the states associated with the irreducible
representations can be used as a basis of zero-brane states.
This is realized by equivariant K-theory since 
$\hbox{K}_{\Gamma}(C^3)\simeq R(\Gamma)$, where $R(\Gamma)$ is the
representation ring of $\Gamma$.
The CFT argument goes through in the presence of discrete
torsion. That is, any zero-brane state has a unique decomposition in
terms of the states associated with the irreducible projective
representations. The question is if there are as many massless
Ramond-Ramond one-form fields as irreducible projective
representations, so that one can associate a generator of the charge
lattice to each irreducible representation. This a priori seems
non-trivial since the projection in the twisted sector is
different\foot{See section $2$ 
for more details.} 
in the presence of discrete torsion and generically the projection
removes these massless fields. The matching between massless
Ramond-Ramond fields and irreducible projective representations follows
in an interesting way from the algebraic properties of discrete
torsion. It turns out that the number of
irreducible projective representations of $\Gamma$ equals
the number of $c$-regular\foot{A group element $g_i\in \Gamma$ 
is $c$-regular if
$c(g_i,g_j)=c(g_j,g_i)\ \forall g_j\in \Gamma$(we take $\Gamma$
abelian for simplicity), 
where $c$ is a cocycle determining the
projective representations $\gamma(g_i)\gamma(g_j)=c(g_i,g_j)\gamma(g_ig_j)$
and the discrete torsion phase
$\epsilon(g_i,g_j)=c(g_i,g_j)/c(g_j,g_i)$. $\epsilon =1$ for
$c$-regular elements. See sections $2$ and $3$ for more
details.} elements of $\Gamma$
\nref\karp{G. Karpilowsky, ``Projective Representations of Finite
Groups'', M. Dekker 1985.}%
\karp . Moreover, the closed string spectrum in the twisted sectors
associated with $c$-regular elements is identical$^{10}$ to the corresponding
twisted sector 
spectrum in the conventional orbifold (without discrete torsion) which
do have massless Ramond-Ramond fields. Thus, each irreducible
projective representation is associated with a generator in the charge
lattice even when there is non-trivial discrete torsion.
This
intuitive result, which follows from algebraic properties of cocycles,
 ties in a nice way the effects of discrete torsion on
open and closed strings. From this CFT result, one is naturally led to 
conjecture that the K-theory of twisted cross products
$K_\Gamma^{[c]}(C^3)\simeq R_{[c]}(\Gamma)$, where 
now $R_{[c]}(\Gamma)$ denotes the module
of projective representations of $\Gamma$ with cocycle $c$.

The organization of the rest of the paper is the following. In section
$2$ we explain the inclusion of discrete torsion in closed string
orbifolds and relate its properties to the topology of the orbifold
group $\Gamma$. We analyze D-branes in these orbifolds and derive
from worldsheet consistency conditions the necessity to use
projective representations when analyzing open strings in orbifolds
with discrete torsion. In section $3$ we find the orbifold analog of
the result by Freed and Witten \freedwitten , present several examples
and describe the charges of fractional branes in orbifolds with
discrete torsion.

\newsec{Open and Closed strings in Orbifolds with Discrete Torsion}

The dynamics of a D-brane at an orbifold singularity
provides a simple example of how the geometry of space-time is encoded
in the D-brane worldvolume theory (for a  partial list of references
see 
\nref\dougmoo{M.R. Douglas and G. Moore, ``D-branes, Quivers, and ALE
Instantons'',  hep-th/9603167.}%
\nref\polch{J. Polchinski, ``Tensors from K3 Orientifolds'',
Phys. Rev. {\bf D55} (1997) 6423, hep-th/9606165.}%
\nref\dkps{M.R. Douglas, D. Kabat, P. Pouliot and S.H. Shenker,
``D-branes and Short Distances in String Theory'', Nucl. Phys. {\bf
B485} (1997) 85, hep-th/9608024.}%
\nref\johmye{C.V. Johnson and R.C. Myers, ``Aspects of Type IIB Theory
on ALE Spaces'', Phys. Rev. {\bf D55} (1997) 6382, hep-th/9610140.}%
\nref\dgm{M.R. Douglas, B.R. Greene and D.R. Morrison, ``Orbifold
Resolution by D-Branes'', Nucl. Phys. {\bf B506} (1997) 84,
hep-th/9704151.}%
\nref\moha{K. Mohri, ``D-Branes and Quotient Singularities of
Calabi-Yau Fourfolds'', Nucl. Phys. {\bf B521} (1998) 161,
hep-th/9707012.}%
\nref\gomdic{D.E. Diaconescu and J. Gomis, ``Duality in Matrix Theory
and Three Dimensional Mirror Symmetry'', Nucl. Phys. {\bf B517} (1998)
53, hep-th/9707019.}%
\nref\mohb{K. Mohri, ``K\"ahler Moduli Space of a D-Brane at Orbifold
Singularities'', Commun. Math. Phys. {\bf 202} (1999) 669, hep-th/9806052.}%
\nref\gre{B.R. Greene, ``D-Brane Topology Changing Transitions'',
Nucl. Phys. {\bf B525} (1998) 284, hep-th/9711124.}%
\nref\mk{S. Mukhopadhyay and K. Ray, ``Conifolds From D-branes'',
Phys. Lett. {\bf B423} (1998) 247, hep-th/9711131.}%
\nref\douglasa{M.R. Douglas, ``Topics in D-geometry'',  hep-th/9910170.}%
 $[31\hskip-3pt-\hskip-3pt 41]$ and for 
interesting applications to AdS/CFT see 
\nref\evasha{S. Kachru and E. Silverstein, ``4d Conformal Field
Theories and Strings on Orbifolds'', Phys. Rev. Lett. {\bf 80} (1998)
4855,  hep-th/9802183.}%
\nref\lnv{A. Lawrence, N. Nekrasov and C. Vafa, ``On Conformal
Theories in Four Dimensions'', Nucl. Phys. {\bf B533} (1998) 199,
hep-th/9803015.}%
\evasha\lnv ).
The low energy gauge theory on
the brane is found by quantizing both open and closed strings on the
orbifold \dougmoo . Closed string modes appear 
as parameters in the gauge theory such 
as in Fayet-Iliopoulos terms and in the superpotential. Open string
modes  provide gauge
fields and scalars which describe the fluctuations of the brane. In
this section we will show that consistency of  interactions between open
and closed strings require embedding an appropriate projective 
representation of the orbifold group on the open string Chan-Paton
factors when studying orbifolds with discrete torsion\foot{Recently, 
\nref\miketom{M.R. Douglas and B. Fiol, ``D-branes and Discrete
Torsion II'', hep-th/9903031.}%
\nref\mukray{S. Mukhopadhyay and K. Ray, ``D-branes on Fourfolds with
Discrete Torsion'', hep-th/9909107.}%
\miketom\mukray\leighbe\ have considered the gauge theory on branes on
an orbifold with discrete torsion.}. We will
start by briefly explaining the essentials of discrete torsion and
describing  the closed string spectrum
of these models. This will be  crucial in determining the appropriate
 projection on open strings.

The spectrum of closed strings on an orbifold $X/\Gamma$ -- with
 abelian $\Gamma$ -- is found by
quantizing strings that are closed up to the action of $\Gamma$ and
 projecting onto $\Gamma$ invariant states. When $\Gamma$ is an
abelian group\foot{In this paper we shall consider $\Gamma$ abelian 
only. It is straightforward to generalize to non-abelian groups.},
 one must quantize and project $|\Gamma|$ 
closed strings. This is reflected in the partition function of the
 orbifold by it  having $|\Gamma|^2$
 terms corresponding to all the possible twists along the
$\sigma$ and $\tau$ directions of the worldsheet.
One loop
modular invariance allows each term in the partition function to
be multiplied by a phase
\eqn\part{
Z=\sum_{g_i,g_j\in \Gamma}\epsilon(g_i,g_j) Z_{(g_i,g_j)},}
such that $\epsilon(g_i,g_j)$ is
invariant under an $SL(2,Z)$ transformation\foot{That is
$\epsilon(g_i,g_j)=\epsilon(g_i^ag_j^b,g_i^cg_j^d)$ where
$\left(\matrix{a &b \cr c &d \cr}\right)\in SL(2,Z)$.} and 
$ Z_{(g_i,g_j)}$ is the partition function of a string closed up
to the action of $g_i\in \Gamma$ with an insertion of action of
$g_j\in \Gamma$ in the trace.

As first noted by Vafa 
\vafa ,
modular invariance on higher genus Rieman surfaces together
with factorization of loop amplitudes imposes very severe
restrictions on the allowed phases. Orbifolds models admitting  these
non-trivial phases are usually referred as orbifolds with discrete
torsion. As we shall briefly explain in a moment, whether 
a particular orbifold model admits such a
generalization depends on the topology of the discrete group $\Gamma$.

In \vafa , Vafa showed that $\epsilon$ must
furnish  a one dimensional representation\foot{For
non-abelian $\Gamma$, $\epsilon(g_i,g_j)$ must be a one
dimensional representation of the stabilizer subgroup $N_{g_i}\in
\Gamma$, where $N_{g_i}=\{g_j\in \Gamma, g_ig_j=g_jg_i\}$.}  of $\Gamma$ 
\eqn\reps{
\epsilon(g_i,g_jg_k)=\epsilon(g_i,g_j)\epsilon(g_i,g_k)}
for each $g_i\in\Gamma$. This provides a natural way to take
$\epsilon$ into account when computing the closed string
spectrum. In orbifolds with discrete torsion, the 
spectrum in the $g_i$ twisted sector is obtained by keeping those
states $|s\hskip-3pt>_i$
in the single
string Hilbert space that satisfy
\eqn\project{
g_j\cdot|s\hskip-3pt>_i=\epsilon(g_i,g_j)
|s\hskip-3pt>_i\qquad \forall
g_j\in \Gamma}
States satisfying \project\ transform in a one dimensional
representation of $\Gamma$. In this language, the spectrum of conventional
orbifolds transform in the trivial one-dimensional representation of
$\Gamma$ where $\epsilon\equiv 1$.

Discrete torsion is intimately connected with the topology of $\Gamma$
via \vafa
\eqn\tors{
\epsilon(g_i,g_j)={c(g_i,g_j)\over c(g_j,g_i)}.}
Here $c\in U(1)$ is a two-cocycle, which is a collection of
$|\Gamma|^2$ phases satisfying the following $|\Gamma|^3$ relations
\eqn\cocycle{
c(g_i,g_jg_k)c(g_j,g_k)=c(g_ig_j,g_k)c(g_i,g_j),
 \qquad \forall g_i,g_j,g_k \in \Gamma.}
  The set of
cocycles can be split into conjugacy classes via the following equivalence
relation compatible with \cocycle
\eqn\eqv{
c^{\prime}(g_i,g_j)={c_ic_j\over c_{ij}}
c(g_i,g_j).}
One can show from the definition of discrete torsion in
\tors\ that indeed $\epsilon$ is a one dimensional representation\foot{One 
can also show that $\epsilon(g_i,g_i)=1$ and
$\epsilon(g_i,g_j)\epsilon(g_j,g_i)=1$.} of
$\Gamma$. Moreover, 
the discrete torsion phase \tors\ is the same for
cocycles in the same conjugacy class. Therefore, the number of
different orbifold models that one can construct is given by the
number of conjugacy classes of cocycles. Topologically, equivalence
classes of cocycles of $\Gamma$ are determined by its second
cohomology \foot{The map $c :\overbrace{\Gamma\times
\Gamma\ldots \times\Gamma}^n\longrightarrow U(1)$ is an $n$-cochain. The set of
all $n$-cochains forms an abelian group $C^n(\Gamma,U(1))$ under
multiplication. One can construct a coboundary operator
$d_{n+1}:C^n(\Gamma,U(1))\longrightarrow C^{n+1}(\Gamma,U(1))$ such
that $d_{n}\circ d_{n+1}=0$ and  write down a corresponding complex.
One can also define the group $Z^n(\Gamma,U(1))=\hbox{Ker} 
d_{n+1}$ of $n$-cocycles
and $B^n(\Gamma,U(1))=\hbox{Im} d_{n}$ of $n$-coboundaries. The
$n$-th cohomology group is defined as usual as $H^n(\Gamma,U(1))=\hbox{Ker}
d_{n+1}/\hbox{Im} d_{n}$. For $n=2$, a $2$-cochain satisfying \cocycle\ maps
to the identity under $d_3$, so it is a 2-cocycle. Moreover,
$d_2\circ c(g_i,g_j)=c_ic_j/c_{ij}$ so the 
equivalence classes of cocycles with \eqv\ as an equivalence relation 
 is given by
$H^2(\Gamma,U(1))$. Cocycles in the same conjugacy class are therefore
cohomologous.}
 group $H^2(\Gamma,U(1))$. Summarizing, given a discrete group
$\Gamma$ there are as many possibly different orbifold models that one
can construct as the number of elements in $H^2(\Gamma,U(1))$.

Placing D-branes in these vacua requires analyzing both open and
closed strings in the orbifold. The closed string spectrum was
summarized in the last few paragraphs in a language that will be
convenient when considering open strings.
 The most general action on an open string is obtained by
letting $\Gamma$ act both on the interior on the string (the
oscillators) and its end-points (the Chan-Paton factors). The open
string spectrum is found by keeping all those states invariant under
the combined action of $\Gamma$
\nref\gp{E.G. Gimon and J. Polchinski, ``Consistency Conditions for
Orientifolds and D-Manifolds'', Phys. Rev. {\bf D54} (1996) 1667,
hep-th/9601038.}%
\gp\dougmoo
\eqn\open{
|s,ab\hskip-3pt>=\gamma(g_i)^{-1}_{aa^{\prime}}|g_i\cdot
 s,a^{\prime}b^{\prime}\hskip-3pt>\gamma(g_i)_{b^{\prime}b},} 
where $s$ is an oscillator state and $ab$ is a Chan-Paton state.
Consistent action on the open string state
and completeness of Chan-Paton wavefunctions demand 
 $\Gamma$ to be embedded on Chan-Paton factors by matrices
the represent $\Gamma$ up to a phase
\eqn\reps{
\gamma(g_i)\gamma(g_j)\propto \gamma(g_ig_j).}
This seems lo leave some arbitrariness since $\Gamma$ may have
several classes of representations. As we will show shortly the
arbitrariness is removed once closed strings are also taken into account.
In particular $\Gamma$ may have several
classes of projective representations where group multiplication is
realized only up to a phase. The
most general such representation is given by
\eqn\general{
\gamma(g_i)\gamma(g_j)=c(g_i,g_j) \gamma(g_ig_j),}
where $c\in U(1)$.
Associativity of matrix multiplication forces $c$ to satisfy the
cocycle condition \cocycle . Moreover, if $c$ satisfies \cocycle ,
so does $c^{\prime}$ defined in \eqv . The corresponding
representation is trivially found to be
$\gamma^{\prime}(g_i)=c_i\gamma(g_i)$.
 Therefore, the different
classes of projective representations of $\Gamma$ are also measured by
$H^2(\Gamma,U(1))$.
Moreover,
the invariant open string spectrum
\open\ of the orbifold model only depends on the cohomology class of the
cocycle and not on the particular representative one chooses.
 This is complete analogy with the closed string discussion
indicating that projective representations should be used when
describing orbifolds with discrete torsion.

We will now show that once we make
 a particular choice of discrete torsion $\epsilon$
in \project\ for the closed strings, that the action on the open
string Chan-Paton factors is uniquely determined to be a projective 
representation \general\ with cocycle $c$. This follows from a
 worldsheet CFT condition demanding $\Gamma$ to be a symmetry of the OPE.
 The action of $\Gamma$ on open and closed
 strings is consistent only if $\Gamma$ is conserved by
 interactions. We already know that this is the case for interactions
 involving only closed strings. 
One must also demand consistency of
 open-closed string interactions, that is $\Gamma$ has to be conserved
 by a open-closed string amplitude\foot{A similar restriction  was
 imposed by Polchinski \polch\ in orientifold models.}.
Let us consider for concreteness the transition between a
 Ramond-Ramond closed 
 string state in the $g_i$-th twisted sector and photon arising from
the open string ending on a D-brane transverse to the orbifold. To
 lowest order in the string coupling this amplitude arises in the disk.
 The closed string vertex operator is built out of
a twist field which creates a cut from its location inside the disk 
to the boundary of
the disk. Fields jump across the cut by the orbifold action $g_i$,
which includes the action of  $\gamma(g_i)$  on the Chan-Paton matrix
$\lambda$ of the open string gauge field. This amplitude
 is completely determined by Lorentz
invariance 
\eqn\ampli{
\hbox{tr}(\gamma(g_i)\lambda)<V_{\alpha}^{i}(0)\widetilde{V_{\beta}^{i}(0)}
V^{\mu}(1)>,} 
where $V_{\alpha}^{i}, \widetilde{V_{\beta}^{i}}$ are the right and
left moving parts of the $g_i$-th twisted Ramond-Ramond
 vertex operator and $V^{\mu}$ is
the vertex operator for the photon. Consistency requires 
invariance of this amplitude under the action of $\Gamma$. As
mentioned earlier, the model is not specified until we choose a
particular discrete torsion $\epsilon$ on the closed string. Therefore, 
taking into
account how $\Gamma$ acts on closed string states \project\ and the
usual adjoint action \open\ on open strings Chan-Paton factors, the
amplitude \ampli\ transforms under the action of $g_j$ as
\eqn\amplit{
\hbox{tr}(\gamma(g_i)\gamma(g_j)^{-1}\lambda\gamma(g_j))
\epsilon(g_i,g_j) 
<V_{\alpha}^{i}(0)\widetilde{V_{\beta}^{i}(0)}
V^{\mu}(1)>.}
Invariance under $\Gamma$ requires setting equal \ampli\ and \amplit ,
which gives after writing the discrete torsion phase in terms of
cocycles
the following constraint
\eqn\proj{
\gamma(g_i)\gamma(g_j)c(g_j,g_i)=c(g_i,g_j)\gamma(g_j)\gamma(g_i).}
This constraint is satisfied by choosing a projective representation
\general . Summarizing, we have shown from simple worldsheet
principles that given an orbifold with discrete torsion \tors\ that
the action of the orbifold group on open strings is determined by the
corresponding cocycle\foot{There seems to be some arbitrariness in
the projective representation one chooses. The worldsheet consistency
condition only determines the cohomology class of the cocycle but does
not pick a particular representative. This freedom, however, does not
affect the spectrum of the orbifold model.}.

It is interesting to note that constraints on $\epsilon$ arise from a
two-loop effect on closed strings but that consistent open-closed
string interactions at tree level determine the action on the open strings.

\newsec{Examples and D-brane Charges}

In this section we will work out a general class of examples and
develop some of the relevant properties of projective representations
of discrete groups that are needed to show the 
results anticipated in section $1$ and $2$. As mentioned in
section $2$ a discrete group $\Gamma$ admits projective
representations if $H^2(\Gamma ,U(1))$ is non-trivial. 
The simplest abelian
group admitting non-trivial projective representations -- or
equivalently, giving rise to discrete torsion in orbifolds -- is
$\Gamma=Z_n\times Z_{n^{\prime}}$. The allowed classes of
representations are labeled by $H^2(\Gamma ,U(1))\simeq Z_d$, where
 $d=gcd(n,n^{\prime})$ is the
greatest common divisor of $n$ and $n^{\prime}$. Thus, a priori, there
are $d$ different orbifold models one can define.

A basic definition in the theory of projective representations
is that of a $c$-regular element. The number of irreducible projective
representations of $\Gamma$ with cocycle $c$ equals the number of
$c$-regular elements of $\Gamma$ \karp\foot{This is very different
to the case of vector representations, for which there are as many
irreducible representations as there are conjugacy classes in the
discrete group.}. A group element $g_i\in \Gamma$
-- for abelian $\Gamma$ -- is
$c$-regular if
\eqn\creg{
c(g_i,g_j)=c(g_j,g_i)\qquad \forall g_j\in \Gamma .}
This definition is independent of the representative of the cocycle
class. Thus, the number $N_c$ of irreducible projective representations with
cocycle class $c$ is given by the following formula\foot{We use the
fact that any non-trivial one-dimensional representation yields zero
when one sums over all group elements.}
\eqn\numbero{
N_c={1\over |\Gamma|}\sum_{g_i,g_j\in \Gamma}{c(g_i,g_j)\over c(g_j,g_i)}=
{1\over |\Gamma|}\sum_{g_i,g_j\in \Gamma}\epsilon(g_i,g_j).}
We have used \tors\ to write the above formula in terms of the discrete
torsion phases. It is clear then that the closed string
spectrum for the sectors twisted by $c$-regular elements is the same
as for conventional orbifolds, so that we have as many irreducible
representations as massless Ramond-Ramond fields of a given rank. As
explained in the section $1$, this prediction should follow from the
algebraic K-theory group  $\hbox{K}_\Gamma^{[c]}(X)$ of twisted cross products.

Let's consider in some detail the example $Z_n\times Z_{n^{\prime}}$.
The discrete torsion phases appearing in the closed string partition
function correspond to one dimensional representations of  $Z_n\times
Z_{n^{\prime}}$. If we let $g_1$ be the generator of $Z_n$ and $g_2$
the generator of $Z_{n^{\prime}}$ a general group element can be
written as $g_1^ag_2^b$, where $a$ and $b$ are integers. 
Then, the allowed discrete torsion
phases are
\eqn\pha{
\epsilon(ab,a^{\prime}b^{\prime})=
\alpha^{m(ab^{\prime}-a^{\prime}b)}, \qquad 
m=0,\ldots,d-1 ,}
where $\alpha=\exp(2\pi i/d)$ and $d=gcd(n,n^{\prime})$. As expected
there are $d$ different phases one can associate to the closed string
partition function \part .

The study of D-branes in these backgrounds require analyzing the
representation theory\foot{This example has also been considered
recently by \leighbe .}%
 of $Z_n\times Z_{n^{\prime}}$. For our purposes,
we only need to find the number of irreducible representations in a
given cohomology class and their  dimensionality. We can use \numbero\
and \pha\ to find how many of them there are. Let $p$ be the
smallest non-zero integer such that 
\eqn\lest{
\exp({2\pi imp\over d})=1,}
then the sum \numbero\  can be split  into sums of blocks of $p$
elements. Usual vector representations correspond to $p=1$. 
If we perform the sum over say $a$ and $b$ for each block we
get $0$ except when  $a^{\prime}$ and $b^{\prime}$ are multiples of $p$,
for which the sum over $a$ and $b$ over a block of $p$-elements just
gives $p^2$. Since there are ${n\over p}$ and ${n^{\prime}\over p}$
blocks of $p$ elements for the sum over $a$ and $b$ and $a^{\prime}$
and $b^{\prime}$ respectively, the total sum yields
\eqn\answ{
N_c={1\over nn^{\prime}}\left({nn^{\prime}\over p^2}\right)^2
p^2={nn^{\prime}\over p^2}.}
Therefore, there are $N_c=nn^{\prime}/p^2$ irreducible projective
representations with cocycle $c$ for $Z_n\times Z_{n^{\prime}}$.
The dimensionality of each irreducible representation can be obtained
from the fact that the regular representation can be decomposed in
terms irreducible ones 
\eqn\rego{
|\Gamma|=\sum_{a=1}^{N_c}d_{R_a}^2,}
where $R_a$ labels the different irreducible representations.
Thus, each representation is $p$-dimensional. Usual vector
representations $(c\equiv 1)$ are one-dimensional and  all irreducible
projective one are bigger.

The conclusion stating that the minimal charge of a D-brane
configuration 
wrapping the entire compact orbifold is an integer bigger than the
minimal charge whenever discrete torsion is non-trivial can be verified by a
simple disk amplitude. We want to compute the charge under the
untwisted sector Ramond-Ramond field corresponding to a wrapped
D6-brane. This can be computed by inserting the untwisted six-brane
Ramond-Ramond vertex operator on the disk. We will sketch the
computation and refer to \fract\ for more details.
The vertex operator has to
be in the $(-3/2,-1/2)$ picture to soak the background superghost charge on
the disk. In this picture and the Ramond-Ramond potential
$C_{\mu_0\ldots\mu_6}$ appears in the vertex operator.
 The amplitude is multiplied by the trace of the
representation acting on Chan-Paton factors for the identity element 
 (to compute the charge under the $g_i$ Ramond-Ramond field one
multiplies by the trace of the representation for $g_i$). The
computation can be easily computed by conformally mapping onto the
upper half plane and imposing the appropriate boundary conditions. The
final result is \fract
\eqn\char{
Q={d_R\over |\Gamma|},}
where $d_R$ is the dimension of the representation considered. This
formula applies both for conventional orbifolds as well as for
orbifolds with discrete torsion.
In the first case one must use
projective representations and in the second vector
representations. Since the smallest irreducible vector representations
of $\Gamma$ 
is one-dimensional but the smallest irreducible projective
representation is larger, this shows that indeed the minimal D-brane
charge allowed for orbifolds with discrete torsion are bigger than for
conventional orbifolds.

\centerline{\bf Acknowledgments}

I would like to express my gratitude to D.E. Diaconescu, M. Douglas, 
E. Gimon, S. Gukov and E. Witten for enlightening discussions. J.G. is
supported in part by the DOE under grant no. DE-FG03-92-ER 40701.

\listrefs

\end